\documentclass[sigconf,9pt,nonacm]{acmart}

\usepackage{xspace}
\usepackage{enumitem}
\usepackage{listings}
\usepackage{tikz}

\newcommand*\circled[1]{\tikz[baseline=(char.base)]{
            \node[shape=circle,draw,inner sep=1pt] (char) {#1};}}

\definecolor{eclipseBlue}{RGB}{42,0.0,255}
\definecolor{eclipseGreen}{RGB}{63,127,95}
\definecolor{eclipsePurple}{RGB}{127,0,85}

\AtBeginDocument{%
  }

\acmConference[Sigmod '25]{International Conference on Management of Data}{June 22--27,
  2025}{Berlin, Germany}

\setlist[itemize]{leftmargin=1.2em, labelsep=0.5em}

\begin{document}

\newcommand{\stitle}[1]{\vspace{0.5ex}\noindent{\bf #1}}
\newcommand{\etitle}[1]{\vspace{0.5ex}\noindent{\em\underline{#1}}}
\newcommand{\eetitle}[1]{\vspace{0.5ex}\noindent{\em{#1}}}

\newcommand{\reffig}[1]{Fig.~\ref{fig:#1}}
\newcommand{\refsec}[1]{Section~\ref{sec:#1}}
\newcommand{\reftab}[1]{Table~\ref{tab:#1}}
\newcommand{\refalg}[1]{Algorithm~\ref{alg:#1}}
\newcommand{\refeq}[1]{Eq.~\ref{eq:#1}}
\newcommand{\refdef}[1]{Definiton~\ref{def:#1}}
\newcommand{\refthm}[1]{Theorem~\ref{thm:#1}}
\newcommand{\reflem}[1]{Lemma~\ref{lem:#1}}
\newcommand{\refex}[1]{Example~\ref{ex:#1}}
\newcommand{\refpro}[1]{Property~\ref{pro:#1}}
\newcommand{\refrem}[1]{Remark~\ref{rem:#1}}

\newcommand{\eat}[1]{}
\newcommand{\todo}[1]{\textcolor{red}{$\Rightarrow$#1}}
\newcommand{\kw}[1]{\textsf{#1}\xspace}
\newcommand{\sys}{\kw{Graphy}}
\newcommand{\fact}{\kw{Fact}}
\newcommand{\dimension}{\kw{Dimension}}
\newcommand{\dimensions}{\kw{Dimensions}}
\newcommand{\problem}{\kw{PDI}}

\newcommand{\scrapper}{\kw{Scrapper}}
\newcommand{\surveyor}{\kw{Surveyor}}

\newcommand{\inspector}{\texttt{Inspection}}
\newcommand{\navigator}{\texttt{Navigation}}
\newcommand{\explorer}{\texttt{Exploration}}
\newcommand{\inductor}{\texttt{Induction}}
\newcommand{\generator}{\texttt{Generation}}
\newcommand{\search}{\texttt{Search}}
\newcommand{\neighborquery}{\texttt{NeighborQuery}}
\newcommand{\statfilter}{\texttt{StatRefiner}}

\title{Graphy'our Data: Towards End-to-End Modeling, Exploring and Generating Report from Raw Data}

\author{Longbin Lai, Changwei Luo, Yunkai Lou, Mingchen Ju$^{\ddag}$, Zhengyi Yang$^{\ddag}$}
\email{{longbin.lailb, pomelo.lcw, louyunkai.lyk}@alibaba-inc.com}
\email{{mingchen.ju@student., zhengyy.yang@}unsw.edu.au}
\affiliation{%
  \institution{Alibaba Group, China; $^{\ddag}$University of New South Wales, Australia}
  \country{}
}
\begin{abstract}
  Large Language Models (LLMs) have recently demonstrated remarkable performance in tasks such as
  Retrieval-Augmented Generation (RAG) and autonomous AI agent workflows. Yet, when faced with large
  sets of unstructured documents requiring progressive exploration, analysis, and synthesis, such as
  conducting literature survey,  existing approaches often fall short. We address this
  challenge -- termed Progressive Document Investigation -- by introducing \sys, an end-to-end platform
  that automates data modeling, exploration and high-quality report generation in a user-friendly manner.
  \sys\ comprises an offline \scrapper that transforms raw documents into a structured
  graph of \fact\ and \dimension\ nodes, and an online \surveyor that enables iterative exploration and
  LLM-driven report generation. We showcase a pre-scrapped graph of over 50,000 papers -- complete
  with their references -- demonstrating how \sys\ facilitates the literature-survey scenario.
  The demonstration video can be found at \url{https://youtu.be/uM4nzkAdGlM}.
\end{abstract}

\maketitle

\section{Introduction}
\label{sec:intro}

We study real-world investigative tasks that require iterative exploration and synthesis of large unstructured data corpora. We refer to this challenge as \textbf{Progressive Document Investigation} (\problem), an iterative process of identifying a focal topic, refining a relevant dataset, and ultimately generating high-quality reports, summaries, or recommendations.
A motivating example of \problem~ is the literature survey process in academic research. Researchers start with a topic of interest, identify a few seed papers, and conduct iterative rounds of investigation: skimming key elements (e.g., ``abstract'', ``challenges'', ``solutions''), following references to additional papers, and expanding the set of relevant works. After collecting a sufficient corpus, they then synthesize their findings into a structured survey report -- often by grouping papers by shared characteristics (e.g., addressing similar challenges or proposing similar solutions).

\eat{
While RAG-based solutions can efficiently retrieve short, context-relevant snippets for question answering, they struggle to maintain consistency and organization when applied to large-scale, multi-step explorations. Existing AI agent systems, on the other hand, risk error propagation across extensive pipelines, especially if they are expected to autonomously parse and link large collections of unstructured data in real time. Moreover, both methods typically provide limited support for iterative user oversight and curation, which researchers often prefer to ensure accuracy and control.
}

The advent of Large language models (LLMs)~\cite{gpt4o} have shown impressive potentials for handling \problem, in particular with techniques like Retrieval-Augmented Generation (RAG)~\cite{lewis2021rag} and autonomous AI agents~\cite{han2024agent}. While these methods excel at single-document queries and conversational workflows, they fall short for solving \problem. 
RAG-based solutions often struggle to maintain consistency and organization when applied to large-scale, multi-step explorations. Existing AI agent systems, on the other hand, risk error propagation across extensive pipelines, especially if they are expected to autonomously parse and link large collections of unstructured data.
Moreover, both methods typically provide limited support for iterative user oversight and curation, which researchers often prefer to ensure accuracy and control.

\begin{figure*}[t]
  \centering
  \includegraphics[width=0.8\linewidth,height=8cm]{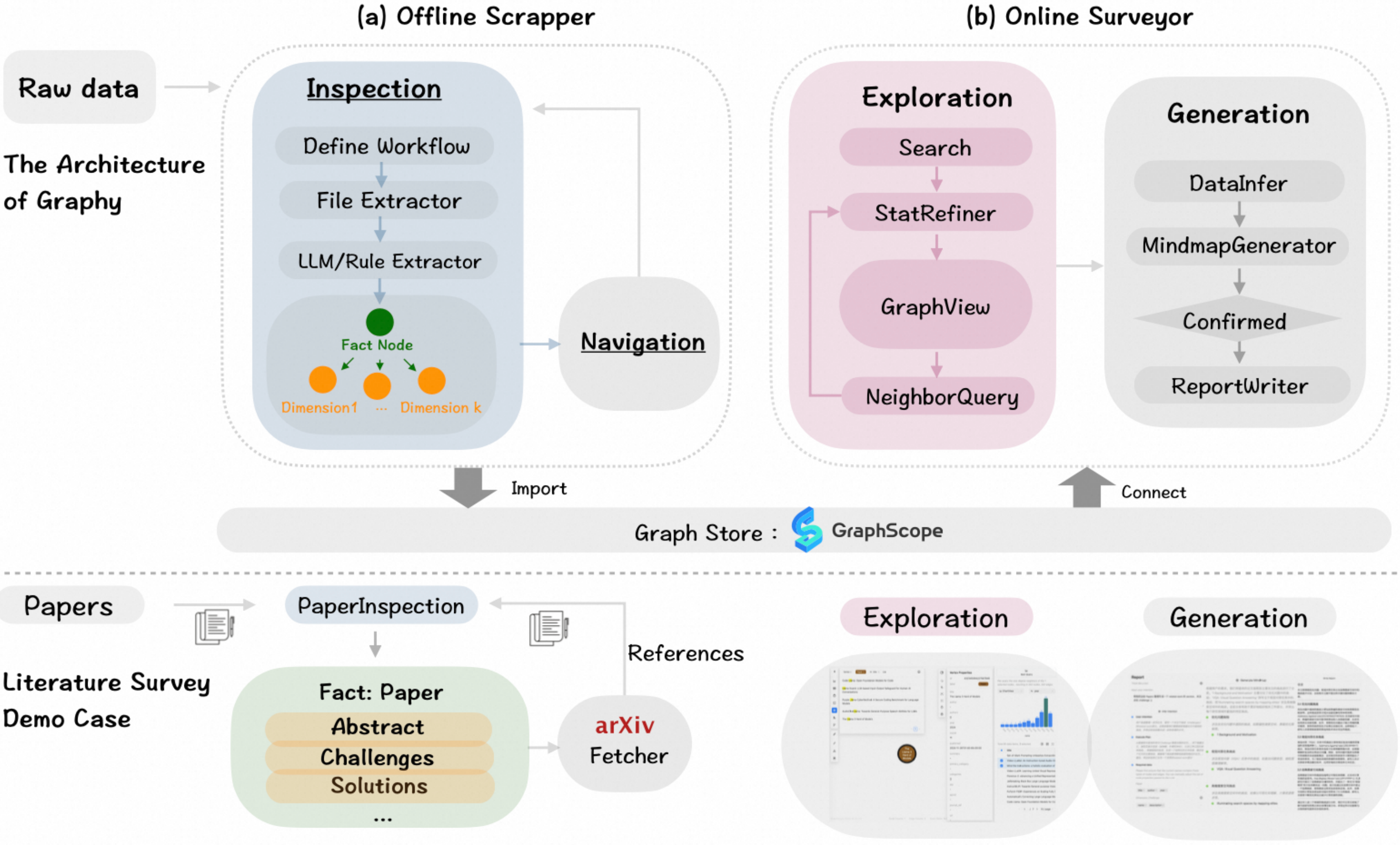}
  \caption{The design and demo case of literature survey of \sys.}
  \label{fig:graphy}
  \vspace*{-1em}
\end{figure*}

To address these gaps, we propose \sys, an end-to-end platform that streamlines the \problem~ workflow.
We adopt the property graph model for the need of iterative exploration in \problem.
Drawing inspiration from business intelligence (BI) systems~\cite{gray1996datacube}, we introduce \fact~ and \dimension~ nodes, analogous to \fact~ and \dimension~ tables in BI. Here, \fact~ nodes represent the primary entities of interest, while \dimension~ nodes capture supplementary information. In a literature-survey context, each paper functions as a \fact~ node (henceforth they are used interchangeably), and its extracted contents, such as ``abstract'', ``challenges'', and ``solutions'', serve as \dimension~ nodes. Although our demonstration centers on the literature-survey scenario, \sys~is broadly applicable; in \refsec{others}, we briefly illustrate its potential in financial use cases.

\reffig{graphy} provides an overview of \sys, which consists of two main roles: the offline \scrapper~ and the online \surveyor.

\stitle{Offline \scrapper}.
The \scrapper allows users to implement the \inspector~ abstraction to direct the extraction of specific
\dimensions~ from each document, often leveraging LLMs.
This step transforms an unstructured document into a structured \fact~ node
linked to predefined \dimension~ nodes. It simulates how a human researcher would skim a document, pinpointing aspects such as abstract, challenges, and solutions. Additionally, a \navigator~ abstraction defines how \fact~ nodes are connected, enabling the retrieval of related items for progressive exploration. For instance, an Arxiv~\cite{arxiv} \navigator~ automatically fetches and downloads research papers from this open-source repository.


Because both the extracted data from the \inspector~ and the linked data from the \navigator~ are relatively stable, we run the \scrapper~ offline. Upon completion, it produces a graph of \fact~ nodes, \dimension~ nodes, and their interconnecting edges, which can be imported into a standard graph database (e.g., GraphScope Interactive~\cite{graphscope}).

\stitle{Online \surveyor}.
Designing a user-friendly \surveyor~ on top of graph databases poses two key challenges. First, unlike SQL, graph query languages are less familiar to users. Second, graph exploration can become unwieldy, particularly with “supernodes,” which have extremely large numbers of connections.  We address these challenges with the \explorer, which is the main interface for navigating the graph and selecting papers of interest. As shown in \reffig{graphy}, it offers a convenient \search~ module to initiate exploration. Users can iteratively move from one set of nodes to their neighbors (referenced papers), with Cypher operations (e.g., \neighborquery) seamlessly integrated into a UI interface inspired by BI toolkits~\cite{polaris}. To avoid overwhelming users, \explorer~ employ histograms and top-k selectors to allow users filter out neighbors of interests.

Eventually, users can proceed to the \generator~ module, which leverages LLMs for creating reports from the papers selected in the \explorer. Users specify the report’s focus (e.g., challenges, solutions), and the \generator~ leverages \fact and relevant \dimension~ nodes to draft a mind map. After reviewing and refining, the system produces a coherent, structured report that mimics a human researcher’s synthesis process. The final document can be exported in various formats (e.g., PDF, LaTeX) to facilitate academic writing.

In this paper, we demonstrate how \sys~ can streamline the literature-survey process. Specifically:

\begin{itemize}
\item \textbf{Data Extraction and Linking:}
With a predefined workflow, we demonstrate how the \scrapper~ employs the \inspector~ to extract \fact~ and \dimension~ nodes from research papers, and how the \navigator~ automatically expands from a set of seed papers to their cited references.

\item \textbf{Paper Exploration:}
Using a pre-scrapped graph containing approximately 50,000 papers, 250,000 \dimension~ nodes, and 160,000 references among the papers, we demonstrate how users can effectively utilize the \explorer~ interface to progressively search for papers of interest, simulating the process of literature survey. 

\item \textbf{Report Generation:}
We demonstrate how the \generator~ collects essential information from the selected papers and creates mind maps in line with users’ intentions. We then showcase its capability to transform these mind maps into a well-structured report, which users can download in formats including PDF and TeX.
\end{itemize}

We have open-sourced both the \sys~ codebase and the pre-scrapped research graph~\cite{opensource}.
The approximate cost of scrapping the research graph using the QWen-plus model~\cite{tongyi} is \$600.

\section{Architecture}
\label{sec:arch}
This section introduces the architecture of \sys, 
which comprises an Offline \scrapper\ 
and an Online \surveyor. 


\subsection{Offline Scrapper}
\label{sec:scrapper}

\stitle{Inspection}. Given a paper document as input, the \inspector\ processes it to produce a structured representation comprising \fact\ and \dimension\ nodes. The paper itself forms the \fact\ node, while its \dimension\ nodes are extracted through a Directed Acyclic Graph (DAG) of instructions. 
Each subnode’s definition aligns with the user’s specific requirements. For simple dimensions (e.g., an ``abstract''), users can employ rule-based methods such as regular expressions. For more advanced tasks, the system supports individually configured LLM subnodes, allowing users to balance cost and performance. For instance, simpler processing can rely on local models~\cite{ollama}, whereas more complex extraction may involve sophisticated cloud-based models~\cite{gpt4o}. These LLM-based subnodes build on a common workflow that chunks PDF text, stores it in a vector database, and then retrieves only the most relevant chunks—based on user-defined queries, for final extraction by the LLM.

\begin{lstlisting}
  "dag": {
      "nodes": [ ...,
        {
          "name": "Abstract",
          "extract_from": { ... },  # the rule of extracting abstract
          "output_schema": { single_typed: ... }  # the output formats
        },
        {
          "name": "Challenges",
          "model" : { "name": "ollama/qwen2.5:7b", ... },
          "query": "Please summarize the challenges in this paper",
          "output_schema": { array_typed: ... }  # the output formats
        },
        {
          "name": "Solutions",
          "model" : { "name": "qwen-plus", ... },
          "query": "Please summarize the solutions in this paper \
              for addressing the above challenges.",
          "output_schema": { array_typed: ... }
        }, ...,
      ],
      "edges": [ ...,

        {"source": "Challenges", "target": "Solutions" }
      ]
  }
  \end{lstlisting}

A snippet of the \texttt{PaperInspection} DAG in \reffig{graphy} is shown above. The subnode \texttt{Abstract} is rule-based for extracting the paper’s ``abstract'', while two LLM-based subnodes form a chain: \texttt{Challenges} uses a locally deployed model (prefixed with “ollama/”) to identify challenges, and \texttt{Solutions} leverages a cloud-based model to extract solutions. Such chain formation allows the \texttt{Solutions} subnode to leverage the context provided by the \texttt{Challenges} subnode.

\stitle{Navigation}. The \navigator\ is responsible for establishing connections between \fact\ nodes,
and in this case particularly, linking papers through their references. Specifically,  a subnode can be
deployed in the above ``PaperInspection'' to extract references from a paper.
These references are then processed by the \navigator\ to fetch additional paper documents. Currently, we have implemented a \navigator\ to retrieve papers from Arxiv~\cite{arxiv}. For each reference, only those that can be matched and retrieved through the \navigator\ are retained. The corresponding documents are downloaded, and the \inspector\ workflow is repeated for these new papers.
In this demo, we only consider the \navigator\ that links papers to the referenced papers. The workflow is actually customizable. For instance, one could implement a \navigator\ to link papers to their associated GitHub repositories.   


\stitle{Graph Modelling}. The results of \inspector\ and \navigator\, as shown in \reffig{graphy}, naturally form a graph comprising \fact\ and \dimension\ nodes. Each \fact\ node represents a paper, while the outputs generated by subnodes in the \inspector\ form a set of \dimension\ nodes linked to their corresponding \fact\ node.
This graph is incrementally expanded as new papers are processed. Specifically, when a new \fact\ node $p_2$ is added, it is linked to an existing \fact\ node $p_1$ if $p_2$ is retrieved by the \navigator\ based on references extracted from $p_1$.

A notable feature of the \inspector\ is the customizable “output\_schema” field for each subnode in the DAG, which defines the schema (data fields and their data types) for the resulting \dimension\ nodes.
The output can be single-typed, such as abstract and title of the paper, which can be directly stored as attributes of the \fact\ node. Alternatively, array-typed outputs like challenges and solutions can be stored as separate \dimension\ nodes, each sharing the same schema.

\eat{
For now, there are tow types of outputs:
\begin{itemize}
	\item Single-typed outputs: These correspond to a single \dimension\ node, which may contain multiple data fields. Examples include fields such as ``abstract'', or ``metadata'' that contains all the metadata including ``title'' of a paper.
	\item Array-typed outputs: These represent multiple \dimension\ nodes, where each node shares the same schema. For example, a ``challenges'' subnode could produce an array of \dimension\ nodes, each representing a distinct challenge.
\end{itemize}

For Single-typed outputs like ``title'' or ``summary'', the information can be directly stored as properties of the \fact\ node while the property graph model is used, eliminating the need to create separate \dimension\ nodes.
}

\subsection{Online Surveyor}
\label{sec:surveyor}


\stitle{Exploration}.  The \explorer~ component is designed to give users an intuitive way to interact with the graph database while minimizing the learning curve. 

\begin{figure}
    \centering
    \includegraphics[width=1.0\linewidth]{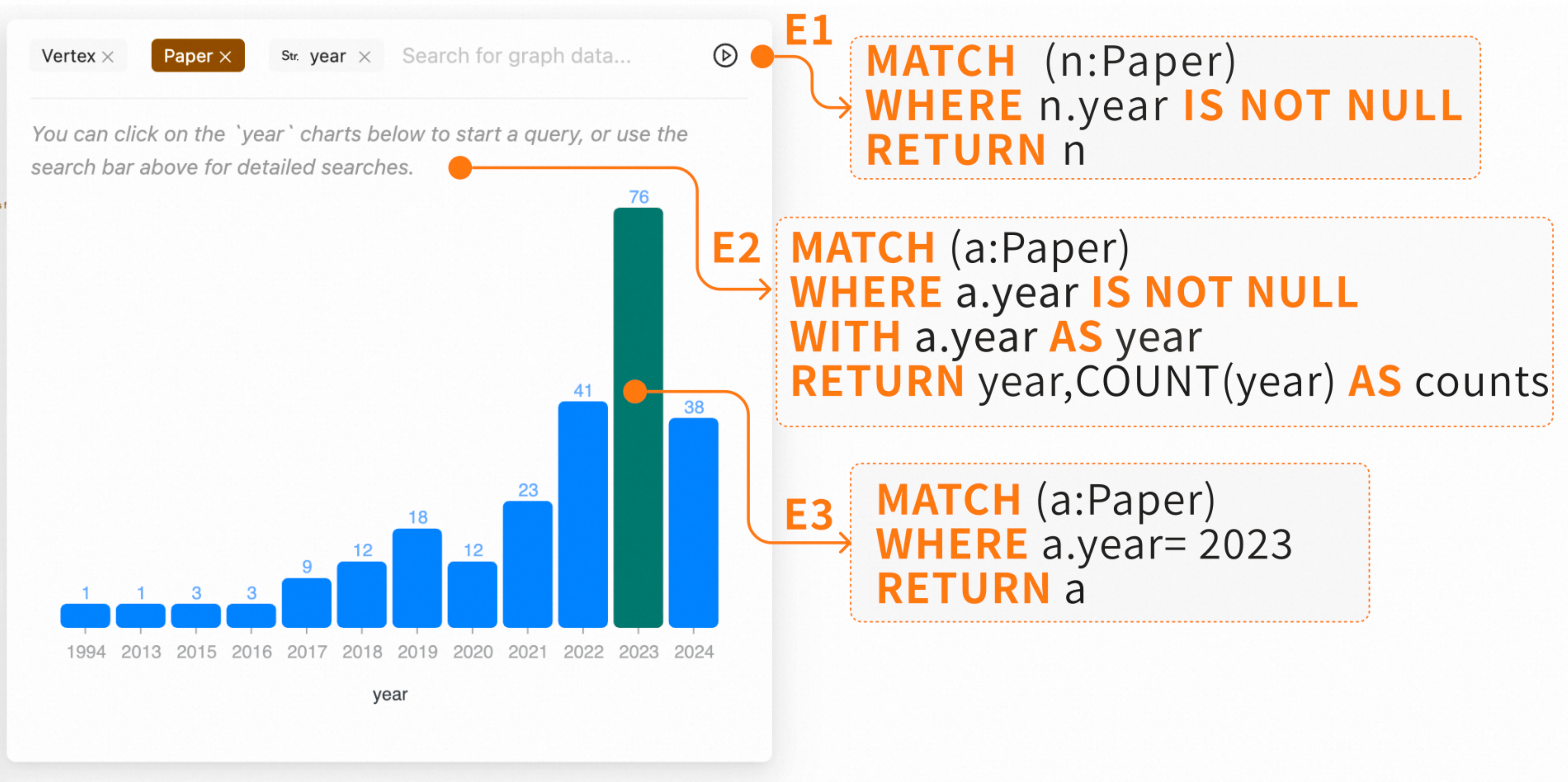}
    \caption{The Search component of \sys.}
    \label{fig:search_component}
    \vspace*{-2em}
\end{figure}

Traditional graph exploration typically relies on query languages, which can require extra effort to master. We address this by embedding graph queries within interactive UI components. As shown in \reffig{search_component}, the Search module in the \explorer~ helps users pinpoint their initial papers for exploration. Three key interactions are highlighted: ``E1'' searches all nodes containing the ``year'' attribute with a single click; ``E2'' displays a histogram of nodes by ``year'' providing a statistical overview; and ``E3'' filters and retrieves nodes for a specific year (e.g., 2023) by clicking the corresponding histogram bar. These user actions are seamlessly translated into Cypher queries and executed on the underlying graph database.

Furthermore, encountering “supernodes” with exceedingly large numbers of connections can often overwhelm users and disrupt the analysis flow. To address this, we introduce a \statfilter~module that intervenes before displaying all the neighbors. This module can present neighbors either as a histogram, allowing users to quickly overview and multi-select by groups, or as a table, where they can sort by specific attributes and choose the top-k results for further exploration. In \refsec{scenario}, we provide examples showing how this approach streamlines the exploration process.

\stitle{Generation.} Once users finish selecting papers in the \explorer, they can employ the \generator~to convert this explored data into structured reports. By leveraging the natural language understanding and summarization capabilities of LLMs, the \generator~ turns the network of interconnected papers on the canvas into a mind map and, ultimately, a well-formatted narrative report. This process involves three main steps: (1) \textbf{Interpreting User Intentions}: Users describe their desired report in natural language, from which LLM infers which attributes and dimensions of the paper are needed. For instance, if a user asks for a related work section focusing on the paper’s challenges, the LLM may determine that the ``title'' and ``abstract'' attributes and the ``challenges'' dimension are required. Users can review and refine these selections before proceeding. As the dimensions are pre-extracted during the offline \scrapper~ phase, the \generator~ can quickly retrieve them on demand.
(2) \textbf{Generating Mind Maps}: Like a human expert, we prompt the LLM to organize the selected papers into a mind map based on the dimensions mentioned by the users, providing a high-level blueprint for the final report. To accommodate context-size limitations, we adopt an iterative approach that feeds the LLM subsets of the data at a time, gradually constructing the mind map for users to review. 
(3) \textbf{Writing Reports}: With the mind map in place, the LLM finalizes the literature survey by generating a cohesive report, which can then be downloaded in various formats (e.g., PDF or TeX) to support academic writing.

\begin{figure*}[t]
  \centering
  \includegraphics[width=0.78\linewidth,height=8cm]{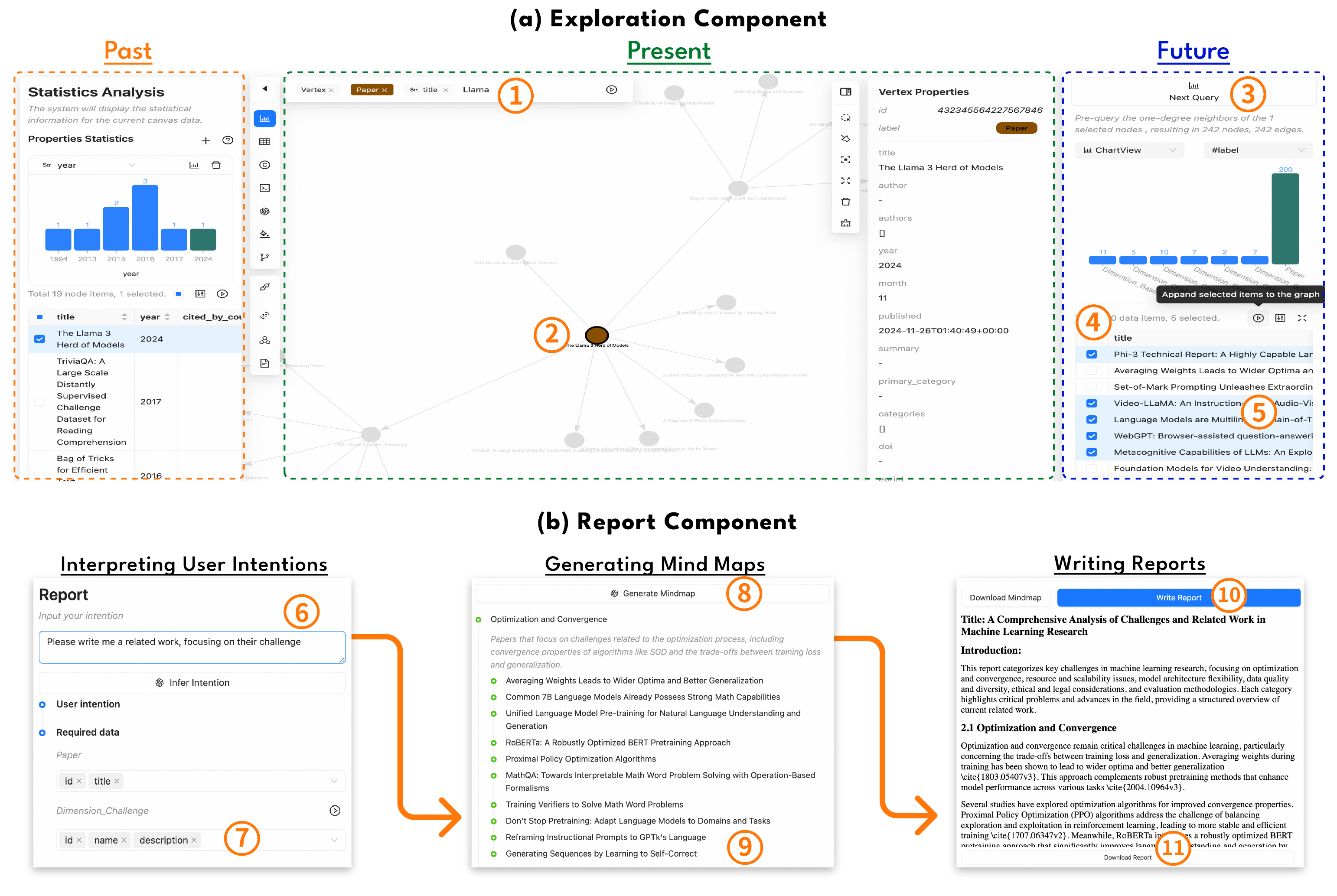}
  \caption{The demonstration scenario of literature survey of \sys.}
  \label{fig:scenario}
  \vspace*{-1em}
\end{figure*}

\section{Demonstrating Literature Survey}
\label{sec:scenario}
We demonstrate how \sys~ applies to literature surveys, with emphasis on the online \surveyor.
We will showcase the functionalities of \scrapper using a video as it is time-consuming. 


The online \surveyor, shown in \reffig{scenario}, allows the demo attendees to explore a pre-extracted paper network containing over 50,000 papers and 160,000 references. We first look into \reffig{scenario}(a) that is the interface of \explorer~  featuring three primary canvases, metaphorically referred to as ``Past'', ``Present'', and ``Future''. Here, ``Past'' displays already explored papers, and ``Present'' shows the currently active papers for reviewing in detail, while ``Future'' highlights the immediate neighbors (i.e., references) of the active papers.
For exploring the papers, the attendee \circled{1} searches for seed papers whose titles contain “Llama3” using the \search~ module; \circled{2} then selects “The Llama 3 Herd of Models” and moves it to the ``Present'' canvas to review its details. Next, \circled{3} the attendee explores the selected paper’s references by pre-querying its neighbors. As described in \refsec{surveyor}, these neighbors are not immediately added to the canvas to avoid overwhelming the user; instead, \circled{4} the \statfilter~ module presents a histogram or table view, allowing attendees to focus on aggregated groups or order the data and finally, \circled{5} decide from the top-k papers for further exploration. By doing so, these papers are added to the ``Present''canvas, while the previously active papers move to the ``Past'' canvas. By iteratively following this workflow, attendees  can explore as many papers as needed, before proceeding to the \generator~task.

In \reffig{scenario}(b), \circled{6} attendees click to input instructions for the report, e.g., ``Please write me a related work, focusing on their challenge''. Based on this input, an LLM (QWen-Plus~\cite{tongyi} for this demo) identifies the relevant attributes and \dimension~ nodes needed for the report, which are \circled{7} displayed for user verification and possible modification. In the example, the LLM highlights the ``Challenge'' node as well as the ``title'' and ``abstract'' attributes from the selected papers. \circled{8} These data are then passed to the LLM to produce a mind map, effectively categorizing the papers according to the identified ``Challenge''.  \circled{9} Attendees can review the mind map, and \circled{10} proceed to final report generation. The final report is built from the mind map and the user’s instructions, culminating in a point-by-point narrative. Once completed, \circled{11} attendees can download the report in PDF or TeX format, complete with citations.

\section{Extension to Financial Scenarios}
\label{sec:others}
We briefly discuss applying \sys\ to two financial scenarios.

\stitle{Company Relationship Analysis}. In this scenario, each company is treated as a \fact~ node,
and the data extracted by \inspector, such as revenues, main business areas and shareholder holdings extracted from financial reports, are represented as \dimension~ nodes.
The \navigator~ component establishes inter-company relationships by leveraging the financial or supply-chain dependencies mentioned in the reports. The generated graph can be used to identify comparable competitors, uncover hidden relationships, or assess contagion effects.
\eat{
Several compelling use cases are enabled:
\begin{itemize}
\item Identifying Comparable Competitors: By clustering or filtering companies with similar profiles (e.g., market segment, growth patterns),
investors can discover promising but less-publicized alternatives to popular stocks. This approach helps avoid the inflated valuations often
found in trending companies and can mitigate investment risk.
\item Assessing Contagion Effects: In the event of a major negative incident (e.g., a significant default or bankruptcy), \sys~
highlights the interconnections that might amplify its impact. For instance, after a large real estate firm files for bankruptcy,
the tool can identify which companies share financial dependencies, supply chain relationships, or shareholder overlap—and therefore
might be exposed to heightened risk.
\end{itemize}
}

\stitle{Financial News Analysis}. In this scenario, each news article serves as a \fact~ node, while pertinent details, such as described events and stock performance indicators, can act as \dimension~ nodes. The \navigator~ builds connections among these \fact~ nodes by identifying shared symbols or overlapping financial metrics. This allows analysts to track the evolution of news stories, assess their market impact, or predict future trends based on historical patterns.

\eat{
\sys~ yields several valuable use cases:
\begin{itemize}
	\item News Clustering and Trend Analysis: Since a single news article may not provide sufficient insight on its own, \sys~ can group related articles to reveal broader patterns. Analysts can thus identify recurring themes or correlated market shifts across multiple sources, improving the reliability of forecasts and decisions.
  \item Assessing the Market Impact of Individual News Items: By referencing historically similar news events and their corresponding market responses, \sys~ enables users to evaluate how sudden developments might affect future market movements. This allows for timely, data-driven judgment on whether a new piece of news is likely to have a major or minor market impact.
\end{itemize}
}

\bibliographystyle{ACM-Reference-Format}
\bibliography{demo}

\end{document}